# Intelligent Metasurface Imager and Recognizer

Lianlin Li[1+*], Ya Shuang[1+], Qian Ma[2], Haoyang Li[1], Hanting Zhao[1], Menglin Wei[1], Che Liu[2], Chenglong Hao[3], Cheng-Wei Qiu[3], and Tie Jun Cui[2*]

[1] State Key Laboratory of Advanced Optical Communication Systems and Networks, Department of Electronics, Peking University, Beijing 100871, China

[2] State Key Laboratory of Millimeter Waves, Southeast University, Nanjing 210096, China

[3] Department of Electrical and Computer Engineering, National University of Singapore, 4 Engineering Drive 3, Singapore 117583

It is ever-increasingly demanded to remotely monitor people in daily life using radio-frequency probing signals. However, conventional systems can hardly be deployed in real-world settings since they typically require objects to either deliberately cooperate or carry a wireless active device or identification tag. To accomplish the complicated successive tasks using a single device in real time, we propose a smart metasurface imager and recognizer simultaneously, empowered by a network of artificial neural networks (ANNs) for adaptively controlling data flow. Here, three ANNs are employed in an integrated hierarchy: transforming measured microwave data into images of whole human body; classifying the specifically designated spots (hand and chest) within the whole image; and recognizing human hand signs instantly at Wi-Fi frequency of 2.4 GHz. Instantaneous in-situ imaging of full scene and adaptive recognition of hand signs and vital signs of multiple non-cooperative people have been experimentally demonstrated. We also show that the proposed intelligent metasurface system work well even when it is passively excited by stray Wi-Fi signals that ubiquitously exist in our daily lives. The reported strategy could open a new avenue for future smart cities, smart homes, human-device interactive interfaces, healthy monitoring, and safety screening free of visual privacy issues.

## Introduction

Internet of Things (IoT) and Cyber Physical Systems (CPS) have opened up possibilities for smart cities and smart homes, which are changing the way people live. In this era, it is highly demanded to remotely probe where people are, what they are doing, what they want to express by their body language, how their physiological states are, etc., in a way not to infringe on visual privacy. Recently developed radio-frequency (RF) sensing technologies have enabled us to realize locating[1-2] and tracking[3-4], notable-action recognition[5-6], human-pose estimation[7-8], breath monitoring[9-10], and others[11-13]. These approaches are preferable since they do not require people to carry any active devices or identification tags. However, these systems are typically designed for one specific task, and hardly perform successive missions adaptively, such as instantly searching for people of interest from a full scene and then adaptively recognizing the subtle body features. Furthermore, they are challenging to monitor the local body gesture language (e.g., hand signs) and vital signs (e.g., respiration and heartbeat) of human beings in real world, because they require people to be deliberately cooperative. Otherwise they involve weak signals that cannot be reliably distinguished from undesirable disturbances. More importantly, these technologies suffer from complicated system designs and extremely expensive hardware cost due to the use of a large number of transmitters and/or receivers to extract the subtle body information. Thus, it is imperative to develop an inexpensive but intelligent device that can instantly obtain the high-resolution image of full human body, instantly focus on an arbitrary local body part of interest, and adaptively recognize the body sign and vital sign in a smart and real-time way. To realize these demands, we propose the concept of ANNs-driven intelligence metasurface for adaptive manipulation of electromagnetic (EM) waves, smart data acquisition, and real-time data processing.

The programmable metasurface, as an emerging active member of metamaterial family[14-23], is an ultrathin planar array of electronically-controlled digital meta-atoms[24-36]. Owing to the unique capability in dynamical and arbitrary manipulations of EM wavefronts, it has elicited many exciting physical phenomena and versatile functional devices, such as the programmable holography[28], computational imager[29-32], wireless communication system[33-35], and others[26, 36]. We here design a large-aperture programmable metasurface for three purposes at one go: 1) to fulfill in-situ high-resolution imaging of multiple people in the full-viewing scene; 2) to rapidly focus the EM fields (including ambient Wi-Fi signals) to chosen local spots and avoid undesired interferences from body trunk and ambient environment; and 3) to monitor the local body signs and vital signs of multiple non-cooperative people in real world by instantly scanning the local body parts of interest.

It is a typical nonlinear EM inverse problem to reconstruct the full-scene image, identify the body language, and monitor the human respiration from acquired measurements in real time, which is a challenging task due to the inherent time-consuming computations and nonunique solutions. It is also not a trivial issue to model and analyze the characteristics of complicated EM environment (e. g. the indoor environment considered in this work) in a tractable way by using conventional approaches. To surpass these difficulties, we propose a cluster of ANNs, three convolutional neural networks (CNNs), for real-time data processing, which can instantly produce the desired results once they are well trained with a large amount of labeled training samples. Due to the easy availability of vast amounts of data and ever-increasing computational power, CNNs have recently been demonstrated a powerful tool in various inverse problems[27-47], including inverse scattering[38-41], metamaterial design[42-44], magnetic resonance imaging[45], and X-ray computation tomography[46]. Our previous results show that the CNN-based strategies can remarkably outperform the traditional techniques in terms of improved reconstruction quality and reduced computational cost[41]. We establish a synergetic network made of three CNNs and implemented with our intelligent metasurface, which are end-to-end mappings from microwave data to the desired images and recognition results. In this way, both global scene and local human body information can be instantly retrieved.

In this article, we present a proof-of-concept intelligent metasurface working around 2.4 GHz (the commodity Wi-Fi frequency) to experimentally demonstrate its capabilities in achieving full-scene images with high resolution and recognizing the human body languages and respirations with high accuracy in smart, real-time and inexpensive way. We experimentally show that our ANNs-driven intelligent metasurface works well under passive stray Wi-Fi signals, in which the programmable metasurface supports adaptive manipulations and smart acquisitions of the stray Wi-Fi signals. Such intelligent metasurface introduces a new way to not only "see" what people are doing, but also "hear" what people talk without deploying any acoustic sensors, even when multiple people are behind obstacles. In this sense, our strategy could bring a new intelligent interface between human and devices, which enables devices to remotely sense and recognize more complicated human behaviors with negligible cost.

## Results

**ANNs-driven intelligent metasurface.** The concept of ANNs-driven intelligent metasurface by integrating the programmable metasurface with deep learning techniques is illustrated in **Fig. 1**. As shown in **Fig. 1(b)**, the designed reflection-type programmable metasurface is composed of 32 × 24 digital meta-atoms with the size of 54×54 mm$^2$, and each meta-atom is integrated with a PIN diode (SMP1345-079LF) for electronic controls. More details on the designed meta-atom and programmable metasurface are elaborated in **Supplementary Figure 1** and **2**. With reference to **Fig. 1(b)**, our intelligent metasurface has active and passive modules of operations. In the active module, the metasurface system includes a transmitter (Tx) to emit the RF signals into the investigation region through Antenna 1, and a receiver (Rx) to detect the echoes bounced from the subject through Antenna 2. In the passive module, the system has two or more coherent receivers to collect the stray Wi-Fi waves bounced from the subject target.

      **Figures 1(c)** illustrates schematically three building blocks of data flow pipeline. In **Fig. 1(c)**, the microwave data collected by the intelligent metasurface are instantly processed with an imaging CNN (the first CNN for intelligent metasurface, called IM-CNN-1for short) to reconstruct the image of whole human body. More details on IM-CNN-1 is given in **Methods** and **Supplementary Figure 3**. Then a well-developed Faster R-CNN[47] is adopted to find the region of interest (ROI) within the whole image, for instance, the chest for respiration monitoring and hand for sign-

language recognition. Afterwards, a modified Gerchberg-Saxton (G-S) algorithm is implemented to come up with the optimal digital coding sequence for controlling the programmable metasurface, so that its radiation wave is focused onto desired spots, as presented in **Supplementary Information.** After receiving the command from the host computer, the programmable metasurface will adaptively focus the EM waves to the desired spots for reading the hand signs or physiological states. As such, not only the unwanted disturbance can be excluded effectively, but also the SNR of echoes from the local body parts of interest can be remarkably enhanced by a factor of 20dB, benefiting the subsequent recognitions of hand signs and vital signs (see **Supplementary Figures 6 and 7**). We develop the other CNN (IM-CNN-2) to process the microwave data to recognize the hand signs. Meanwhile, the human breath is identified by time-frequency analysis of microwave data. More details on IM-CNN-2 and respiration identification algorithm are given in **Supplementary Figures 4**. Several sets of representative results are recorded in **Supplementary Videos 1** and **2.**

**In-situ imaging of whole human body using IM-CNN-1.** We first present the in-situ high-resolution microwave imaging of whole human body in the active mode, which are conducted in our lab environment. In this scenario, the intelligent metasurface system has two horn antennas connected to two ports of Agilent vector network analyzer (VNA). One antenna is used to transmit EM signals into the investigation domain, and the other is to receive the EM echoes bounced from the specimen. In the high-resolution imaging, the programmable metasurface serves as a spatial microwave modulator controlled by the field programmable gate array (FPGA) to register the information of specimen in compressive-sensing manner (see **Supplementary Information**).

The *kernel* of the intelligent metasurface for whole-body imaging is IM-CNN-1 to process the microwave data instantly. To obtain large amount of labeled samples for training IM-CNN-1, a commercial 4-megapixel digital optical camera is embedded in the intelligent metasurface system. The training samples captured by the camera are used to train IM-CNN-1 after being pre-processed with background removal, threshold saturation, and binary-value processing (see **Supplementary**

**Figure 3**). The labeled human-body images can be approximately regarded as EM reflection images of the human body over the undergoing frequencies from 2.4 to 2.5 GHz. We collect $8 \times 10^4$ pairs of labeled training samples in our lab environment, and it takes around 8 hours to train IM-CNN-1. The trained IM-CNN-1 can then be used to instantly produce a high-resolution image of human body in less than 0.01 seconds.

We experimentally characterize the performance of the intelligent metasurface in achieving high-resolution images of the whole human body and simultaneously monitoring notable movements in indoor environment. Two volunteers (coauthors Shuang Ya and Hao Yang Li, referred to as training persons) with different gestures are used to train the intelligent metasurface; while three persons (coauthors Shuang Ya, Hanting Zhao, and Menglin Wei, referred to as testing persons) are invited to test it. The trained intelligent metasurface is then used to produce high-resolution images of the test persons, from which their body gesture information can be readily identified. A series of imaging results are presented in **Figure 2** and **Supplementary Video 1.** Particularly, the "see-through-the-wall" ability is validated by clearly detecting notable movements of the test persons behind a 5cm-thick wooden wall. Selected results are provided in the rightest column of **Figure 2**, where the corresponding optical images and microwave raw data are provided as well. To examine the imaging quality quantitatively, **Supplementary Figure 5(a)** compares the image quality versus the number of random coding patterns of the programmable metasurface in terms of the similarity structure index metric (SSIM)[34]. We show that it is enough to achieve the high-quality images by using 53 coding patterns, where 101 frequency points from 2.4 to 2.5 GHz are utilized for each coding pattern. As reported in **Supplementary Information**, the switch time of coding patterns is around 10μs, implying that the time in data acquisition is less than 0.7ms in total even if 63 coding patterns are used. Consequently, we safely conclude that the intelligent metasurface integrated with IM-CNN-1 can instantly produce high-quality images of multiple persons in real world, even when they are behind obstacles.

**Recognition of hand signs and respirations.** After obtaining the high-resolution image of whole body, the intelligent metasurface is then used to recognize the hand signs and vital signs adaptively in real indoor environments. This capacity is benefitted from the robust feature of the intelligent metasurface in adaptively focusing the EM energy to the desired spots with very high spatial resolution. This feature supports accurate detections of EM echoes reflected from the human hand for recognizing the sign language or from the body chest for identifying respiration. Typically, the rate of the human hand sign language and respiration is in the order of 10~30 bps, which is drastically slower than the switching speed of coding patterns by a factor of $10^5$. Thus the radiation beams of the intelligent metasurface are manipulated to rapidly scan the local body parts of interest in each observation time interval. As a result, we realize to monitor the hand signs and respirations of multiple people simultaneously in a time-division multiplexing way (see **Supplementary Figure 4**).

To fulfil the complicated task, we propose a three-step routine procedure. Firstly, the Faster R-CNN[47] is applied to extract the hand or chest part from the full-scene image obtained with IM-CNN-1 in a divide-and-conquer manner. Secondly, the metasurface is manipulated by adaptively changing its coding pattern to make its radiation beam point to the hand or chest (see **Figures 3a-c**). Thirdly, IM-CNN-2, an end-to-end mapping from the microwave data to the label of hand sign language, is developed to recognize the hand signs. The conventional time-frequency analysis is performed for detecting the respiration (see **Supplementary Figure 4**).

The training samples of IM-CNN-2 include ten hand signs (see **Figure 3(a)**, corresponding to ten different English letters), and 8000 samples for each hand sign. Thus we have 80000 samples in total. **Figure 3(d)** reports the classification matrix for the ten hand signs with average recognition accuracy of above 95% by using the intelligent metasurface integrated with IM-CNN-2, where the test people are behind a 5cm-thickness wooden wall. We clearly see that the performance of hand-sign recognition is almost not effected by the number of test persons after the hand parts are well identified by Faster-R-CNN.

The respiration is an important health metric for tracking human physiological states (e. g. sleep, pulmonology, and cardiology). Similar to the recognition of human hand signs, we use the intelligent metasurface to monitor the human respiration with high accuracy. **Figure 3(e)** reports the monitored results of respiration of two test persons behind the wood wall. We observe that the normal breathing and holding breathing are clearly distinguished, and the respiration rate can further be identified with the accuracy of 95% and beyond, where the ground truth is achieved by a commercial breathing monitoring device. It can be expected that the identification performance is almost independent on the number of test persons due to the use of time-division multiplexing respiration detection.

**Intelligent metasurface with stray Wi-Fi signals.** Our intelligent metasurface works around 2.4-2.5GHz, which is exactly the frequency of commodity Wi-Fi signals. We here investigate the performance of high-resolution imaging of full scene and recognition of human hand signs and vital signs when the metasurface is *excited by the commodity stray Wi-Fi signals*. For simplicity, we particularly consider to use Wi-Fi beacon signals. In this case, the intelligent metasurface works differently in three major aspects. Firstly, the stray non-cooperative Wi-Fi signals are dynamically manipulated by the metasurface. Secondly, two or more coherent receiving antennas are used to acquire the Wi-Fi signals bounced from the subject specimen with the aid of oscilloscope (Agilent$^{TM}$ MSO9404A). Thirdly, the microwave data acquired by receivers are coherently preprocessed before sending to IM-CNN-1 such that the statistical uncertainties on stray Wi-Fi signals can be calibrated out. More details can be found in **Supplementary Video 2** and **Supplementary Information**.

**Figure 4(a)** presents a set of in-situ passive imaging results of the subject person behind the wooden wall in our indoor lab environment, where random coding patterns are also used in the programmable metasurface. We surprisingly note that the imaging results by the commodity stray Wi-Fi signals are comparable to those in the active mode. Based on the high-resolution images of the full human body, we can realize the recognition of hand signs and vital signs by adaptively performing the three-step routine procedure in the active mode. In particular, the Faster R-CNN is operated on the full-scene image to instantly find the location of hand or body chest; then suitable coding patterns of the intelligent metasurface can be achieved and controlled so that the stray Wi-Fi signals are spatially focused and enhanced on the desired spots; and finally IM-CNN-2 or the time-frequency analysis algorithm is used to realize the recognition of hand signs and vital signs. As shown in **Fig. 4(b)-(c)**, the commodity Wi-Fi signals can be well focused at the desired location, e.g., the left hand of the subject person, by using the developed intelligent metasurface. As a result, SNR of Wi-Fi signals can be significantly enhanced with a factor of more than 20dB, which is directly helpful in the subsequent recognitions of hand signs and vital signs (see **Supplementary Figures 6 and 7**). **Figures 4(d)** and **4(e)** show experimental results of the hand-sign and respiration recognitions of two people, which have better accuracies of 90% and 92%, respectively. To summarize, even with the illumination of stray Wi-Fi signals, the proposed intelligent metasurface can realize high-resolution images of full scene and high-accuracy recognitions of hand signs and vital signs of multiple people in a smart and real-time way in the real world.

## Conclusions

We cast the concept of intelligent metasurface imager-cum-recognizer, showing its robust performance in remotely monitoring notable human movements, subtle body gesture languages, and physiological states from multiple non-cooperative people, in real-world settings. The developed ANNs-driven intelligent metasurface relies on two key issues: 1) a large-aperture programmable metasurface for adaptive manipulation of EM wavefields and smart data acquisition; and 2) three ANNs for smart processing of data flow in real time. We further experimentally demonstrated that the intelligent metasurface works well even when it is passively excited by commodity Wi-Fi signals. Such strategy can not only monitor the notable or non-notable movements of non-cooperative persons in a real world, but also help the many handicapped to remotely send commands to devices using body languages. We expect that lip reading and human-mood recognition could also be realized if higher resolution and accuracy are achieved by involving higher frequencies. In principle, the concept of intelligent metasurface can be extended over the entire EM spectra, which will open a new avenue for future smart homes, human-device interactive interfaces, healthy monitoring, and safety screening.

## Methods

**Design of programmable metasurface.** The designed programmable metasurface consists of 32 x 24 meta-atoms operating at around 2.4 GHz, as shown in **Supplementary Figure 1**, in which the details of electronically-controllable meta-atom with size of 54mm×54mm are illustrated in **Supplementary Figure 2**. In each meta-atom, a PIN diode (SMP1345-079LF) is integrated to control its EM reflection phase, and the equivalent circuits of the diode for the ON and OFF states are presented in **Supplementary Information**. The meta-atom is composed of two substrate layers: the top layer is F4B with the relative permittivity of 2.55 and loss tangent of 0.0019; and the bottom is FR4 with size of 0.54 ×0.54 mm$^2$. The top square patch, integrated with a SMP1345-079LF PIN diode, has a size of 0.37× 0.37 mm$^2$. In addition, a Murata LQW04AN10NH00 inductor with inductance of 33nH is used to achieve the good isolation between the RF and DC signals. CST Microwave Studio is used to design the meta-atom: 1) the reflection response of the meta-atom is investigated under different operation states of the PIN diode; 2) a Floquet port is used to produce an *x*-polarized wave incidence on the metasurface and monitor the reflected wave; and 3) periodic boundary conditions are set to the four sides to model an infinite array.

**Configuration of intelligent metasurface.** The intelligent metasurface has two operational modes: active and passive modes. In the active mode, the intelligent system is composed of a large-aperture programmable metasurface, three CNNs, a transmitting antenna, a receiving antenna, and a vector network analyzer. In the passive mode, it includes the programmable metasurface, three CNN networks, a pair of receiving antennas, and an oscilloscope, in which one antenna serves as a reference receiver to calibrate out the undesirable effects from the system error. An optical digital camera is synchronized with the whole intelligent metasurface, which is used to collect the labeled samples to train the deep ANNs.

The large-aperture programmable metasurface is designed to control ambient EM wavefields dynamically and adaptively by using FPGA by manipulating its coding sequences, which has two-fold roles. Firstly, it serves as a relay station of information or an electronically-controllable random mask, transferring the EM signals carrying finer information of specimen to the receivers. Secondly and more importantly, to realize body-language recognition and respiration monitoring, the programmable metasurface with optimized coding patterns can focus the EM wavefields on the desired spots, and meanwhile suppress the irrelevant inferences and clutters.

**IM-CNN-1, IM-CNN-2, and Faster R-CNN.** The intelligent metasurface is configured with three deep CNNs for smart and real-time data processing. IM-CNN-1 is designed for transferring the EM raw data into the image of the whole human body. The Faster R-CNN is a popular classifier originally developed in the area of computer vision[47], and here is used to identify the hand and chest from the reconstructed whole image. IM-CNN-2 is a classifier to infer the human hand signs from the microwave data.

IM-CNN-1 and IM-CNN-2 operate directly on the microwave raw data, in which the training stage is performed by the ADAM optimization method with the mini-batch size of 32 and epoch of 101. The learning rates are set to $10^{-4}$ and $10^{-5}$ for the first two layers and the last layer, and halved once the error plateaus. The complex-valued weights and biases are initialized by random weights with zero-mean Gaussian distribution and standard deviation of $10^{-3}$. The trainings are performed on a workstation with Intel Xeon E5-1620v2 central processing unit, NVIDIA GeForce GTX 1080Ti, and 128GB access memory. The machine learning platform Tensor Flow[48] is used to design and train the networks in the intelligent metasurface system.

## Acknowledgments

This work was supported by the National Key Research and Development Program of China under Grant Nos. 2017YFA0700201, 2017YFA0700202, and 2017YFA0700203, the National Natural Science Foundation of China under Grant Nos. 61471006, 61631007, and 61571117, and the 111 Project under Grant No. 111-2-05.

## Author Contribution

L.L. conceived the idea, conducted the numerical simulations and theoretical analysis. T.J.C. proposed the concept of programmable metasurfaces, and L. L., C.H., C.W.Q, T. J. C. wrote the manuscript. All authors participated in the experiments, data analysis, and read the manuscript.

## Additional Information

**Supplementary Information** accompanies this article.

**Competing Interests**: The authors declare no competing financial interests.

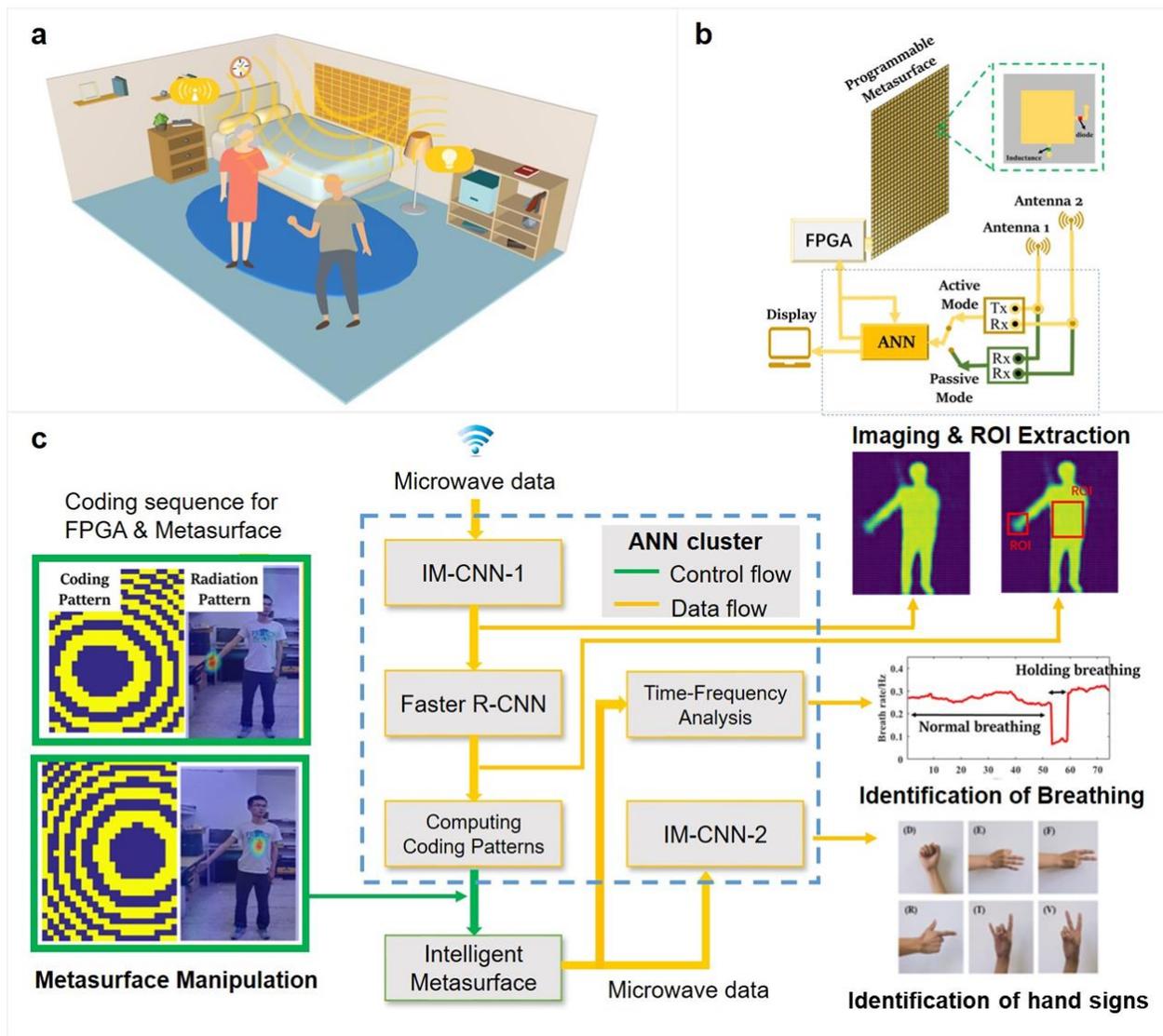

**Figure 1 | Working principle of the intelligent metasurface. a**, an illustrative scenario for monitoring peoples in a typical indoor environment in a smart, real-time and inexpensive way, where the intelligent metasurface decorated as a part of wall is used to adaptively manipulate ambient Wi-Fi signals. **b**, The schematic configuration of intelligent metasurface system by coming a large-aperture programmable metasurface for manipulating and sampling the EM wavefields adaptively with artificial neural networks (ANNs) for controlling and processing the data flow instantly. The intelligent metasurface has two operational modes: active and passive modes. In the active mode, the intelligent system has a transmitting antenna and a receiving antenna. In the passive mode, the intelligent system has a pair of receiving antennas. In addition, the photo of fabricated large-aperture programmable metasurface and the map of meta-atom. (**c**), Microwave data processing flow by using deep learning CNN cluster. In (**c**), the microwave data are processed with IM-CNN-1to form the image of the

whole human body. Then, the Faster R-CNN is performed to find the region of interest (ROI) from the whole image, for instance, the chest for respiration monitoring, and the hand for sign language recognition. Afterwards, the G-S algorithm is used to find the coding sequence for controlling the programmable metasurface such that its associated radiation beams can be focused toward the desirable spot. IM-CNN-2 processes microwave data to recognize the hand sign; and the human breathing is identified by the time-frequency analysis of microwave data.

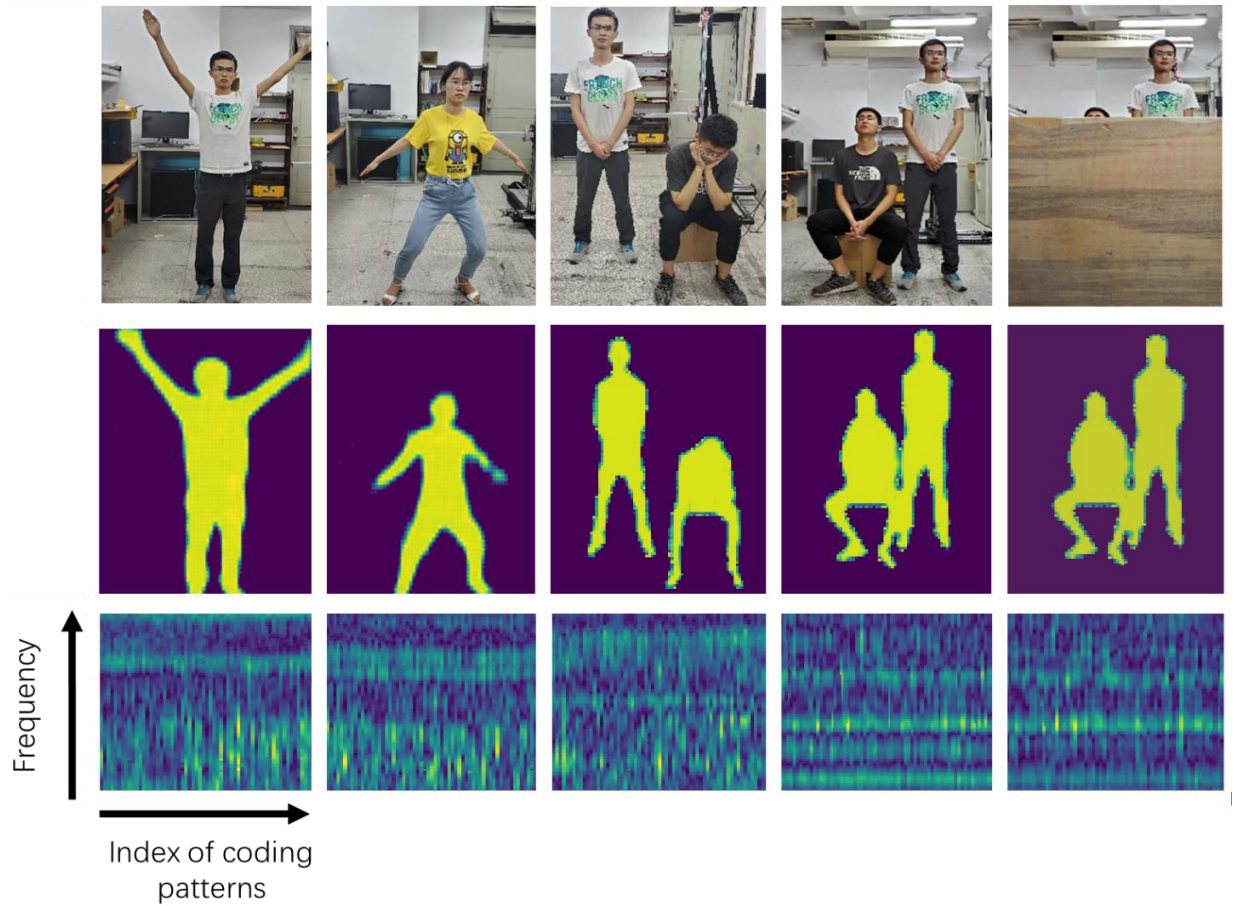

**Figure 2 | In-situ imaging results using the intelligent metasurface with active microwave. (Top row)** The first row shows the optical images of specimen, which include single person with different gestures, two persons with different gestures, and two persons behind a 5cm-thick wooden wall. (**Middle**) The second row illustrates the corresponding imaging results by the intelligent metasurface with IM-CNN-1. (**Bottom**)The bottom row presents corresponding amplitudes of microwave data.

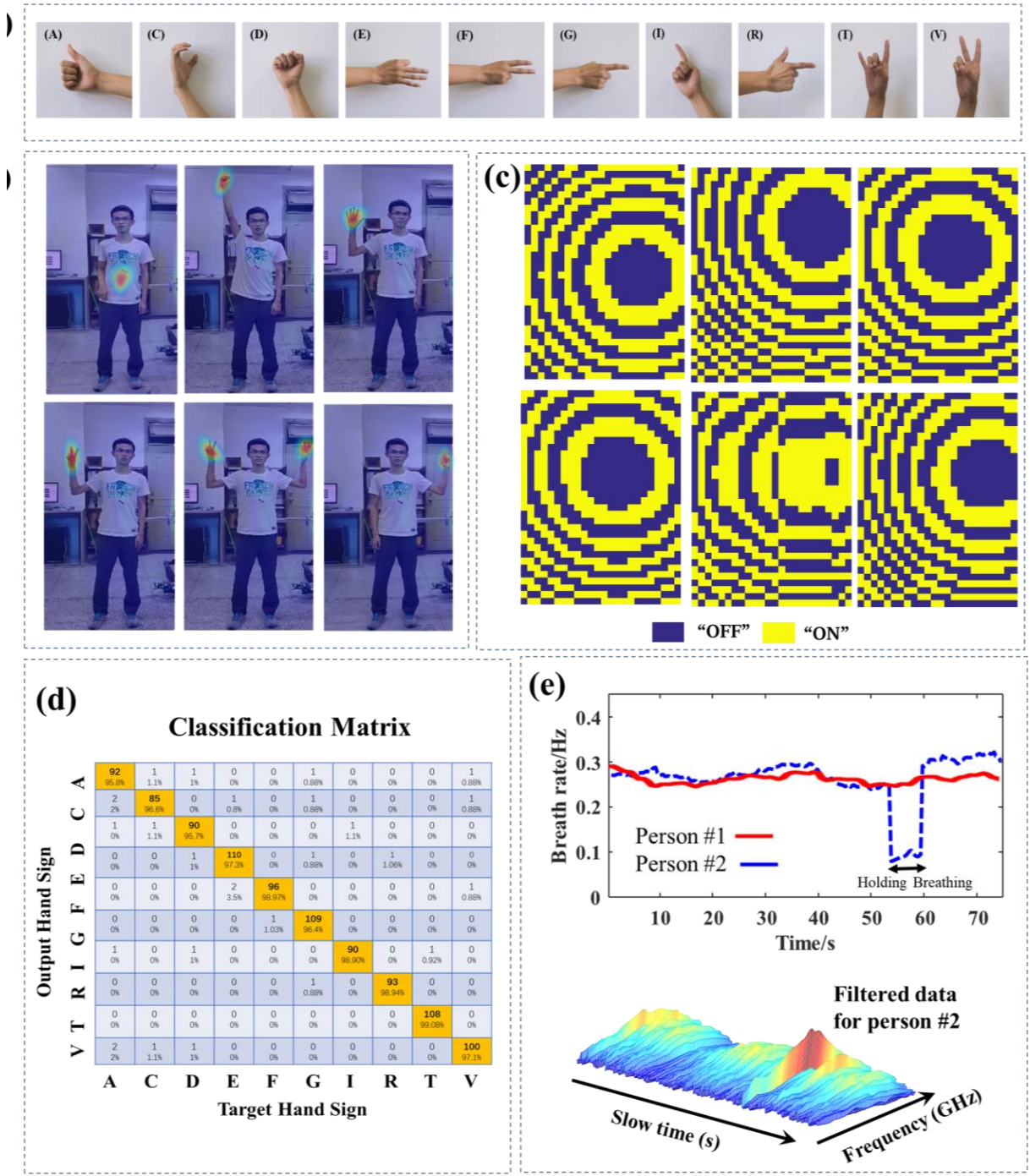

**Figure 3 | Recognition results of human hand signs and respirations by the intelligent metasurface with active microwave.** (**a**) Ten hand signs of English letters considered in this work. (**b**)-(**c**) Selected results of the microwave radiations focused at the desirable spots, for instance, human hands and chest, and corresponding optimized coding patterns of programmable metasurface. In (b), the wavefield distributions are obtained using so-called near-field scanning technique (**see Supplementary Note 5**). (**d**) The classification

matrix of 10 hand signs in (a) obtained by using the IM-CNN-2. **(e)** Results of human respiration of two persons in our lab environment, where person#1 has the normal breathing, and person #2 holds his breathing at around 55s. From this figure, one can clearly see that not only two states of normal breathing and holding breathing can be readily distinguished, but also the respiration can be accurately identified. In addition, the microwave data with motion filter is also provided.

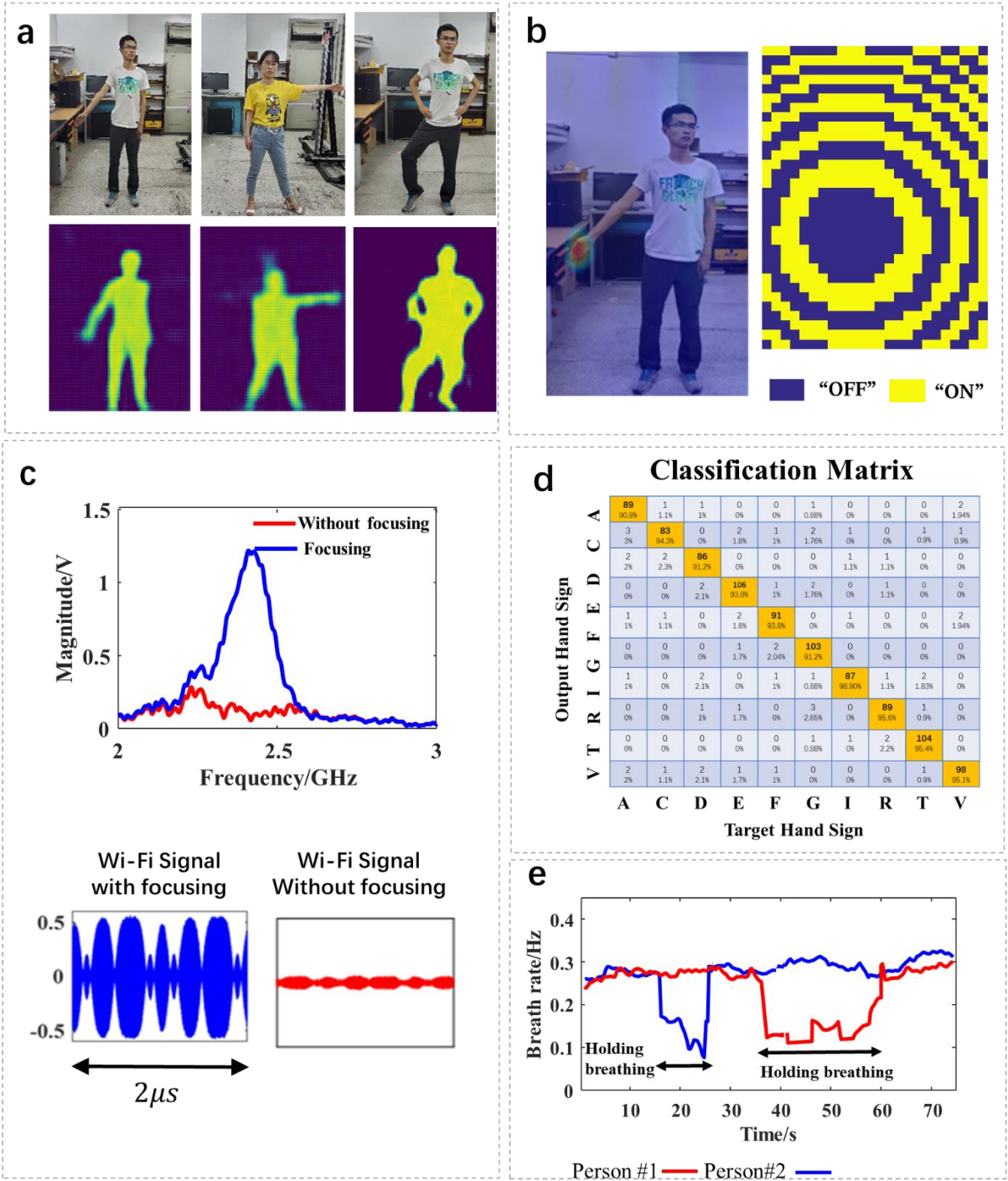

**Figure 4 | Experimental results of in-situ imaging, hand-sign recognition, and respiration identification using the intelligent metasurface in the passive mode with commodity stray Wi-Fi signals.** (**a**) In-situ imaging results using the intelligent metasurface excited with commodity Wi-Fi signals. The first row shows the optical images of the subject person with different gestures behind a 5cm-thick wooden wall. The second row reports corresponding imaging results by the intelligent metasurface with IM-CNN-1. (**b**) On the left is the result of the Wi-Fi signals focused at the desirable spot of human hand, and on the right is corresponding coding pattern of programmable metasurface. Here, the spatial distribution of Wi-Fi signals is obtained using so-called near-field scanning technique, as done in Fig. 3(c). (**c**) The Wi-Fi signals with and without being focused through the programmable metasurface have been compared, which are measured at the location of left hand shown in (b). The top row compares the frequency spectrums of Wi-Fi signals, which are obtained by operating on the raw time-domain Wi-Fi signals with standard FFT technique. Note that the signal-to-noise ratio of Wi-Fi signals at the local spot of human hand can be enhanced by a factor of more than 20dB at around 2.4GHz. (**d**) The classification matrix of 10 hand signs in Fig. 3(a) obtained by using the IM-CNN-2. (**e**) Results of human respiration of two non-cooperative persons behind a 5cm-thickness wall in our lab environment. From this figure, one can clearly see that not only two states of normal breathing and holding breathing can be readily distinguished, but also the respiration can be accurately identified.